\documentclass{llncs}

\usepackage{hyperref}
\usepackage{amsmath}
\usepackage{amsfonts}
\usepackage{pgfplots}
\usepackage{float}
\usepackage{caption}
\renewcommand{\figurename}{F.}
\usepackage{todonotes}

\begin{document}

\title{GOAL-DTU: \\[1ex] Development of Distributed Intelligence for the Multi-Agent Programming Contest}


\author{Alexander Birch Jensen \and J{\o}rgen Villadsen}

\institute{Algorithms, Logic and Graphs Section \\
Department of Applied Mathematics and Computer Science \\
Technical University of Denmark \\
Richard Petersens Plads, Building 324, DK-2800 Kongens Lyngby, Denmark} 
\maketitle

\thispagestyle{plain} \pagestyle{plain}

\smallskip

\begin{abstract}
We provide a brief description of the GOAL-DTU system for the agent contest, including the overall strategy and how the system is designed to apply this strategy. Our agents are implemented using the GOAL programming language. We evaluate the performance of our agents for the contest, and finally also discuss how to improve the system based on analysis of its strengths and weaknesses.
\end{abstract}

\smallskip  

\section{Introduction}

In fall 2019 we participated as the GOAL-DTU team in the annual Multi-Agent Programming Contest (MAPC).
We are using the GOAL agent programming language \cite{GOAL1,GOAL2,GOAL3} and we are affiliated with the Technical University of Denmark (DTU).
We participated in the contest in 2009 and 2010 as the Jason-DTU team \cite{Boss+2010,Vester+2011}, in 2011 and 2012 as the Python-DTU team \cite{Ettienne+2012,Villadsen+2013}, in 2013 and 2014 as the GOAL-DTU team \cite{Villadsen+2013-GOAL}, in 2015/2016 as the Python-DTU team \cite{Villadsen+2017} and in 2017 and 2018 as the Jason-DTU team \cite{Villadsen+2018}.

In 2019 we had the new \textit{Agents Assemble} scenario. The paper is organized as follows:
\begin{itemize}
    \item Section \ref{agentproggoal} describes agent programming using the GOAL language.
    \item Section \ref{strategy} covers the overall strategy of our agents.
    \item Section \ref{agentknowledge} describes the knowledge our agents acquire from the environment.
    \item Section \ref{agentcomm} describes how our agents communicate.
    \item Section \ref{agentmovement} describes the movement of our agents.
    \item Section \ref{solvingtasks} covers how our agents complete selected tasks.
    \item Section \ref{evaluation} evaluates the performance of our agents in the three matchups.
    \item Section \ref{discussion} discusses improvements to the system.
    \item Section \ref{conclusion} is our conclusion.
\end{itemize}

\section{Agent Programming in GOAL}\label{agentproggoal}
This section introduces the basic concepts of the GOAL agent programming language that are relevant to the implementation of our system. Remarkably, the programming philosophy behind GOAL is quite different when compared to other popular agent programming languages. 

Agents in GOAL are to be understood as self-controlled independent entities. Each agent interacts with the environment and communicates with other agents. Percepts and messages are treated as events that can be processed. This event processing then feeds into the knowledge, beliefs and goals (the three components that comprise an agent's cognitive state).

Beyond the cognitive state, the core abilities of an agent are the just mentioned event processing capability, the ability to represent knowledge and reason about it, and finally, rule-based decision-making which allows an agent to select an action based on its current cognitive state.

\subsection{The GOAL Execution Loop}
GOAL features a simple execution loop of each agent. Beyond an initialization module that can process the initial state of the environment, and set up the initial cognitive state, GOAL follows the execution loop below:
\begin{enumerate}
    \item \textbf{Check new events}: If there are no new events, the next step is skipped.
    \item \textbf{Process events}: The event module processes new events. Recall that these events are either percepts from the environment or messages from other agents. It is the purpose of this module to update the cognitive state of the agent before selecting the next action
    \item \textbf{Action selection}: The main module defines the rules for decision-making. Based on the rules, the first valid action is selected. Note that several actions may be applicable based on the rules in the main module. GOAL allows for other strategies, but it is essential to our implementation that the first action is always the one that is selected.
    \item \textbf{Perform action}: The selected action is sent to the environment (and communication actions are executed internally).
\end{enumerate}

Technically speaking, there is also a final step that applies the post-conditions (effects) for the selected action from our action specification. However, it is not relevant for us since we rely solely on the event module to perceive the effects of actions.

\subsection{Action Selection}
GOAL advocates that agents are individual entities that reason about their environment. They react to changes in their environment rather than executing predetermined plans. To exemplify, an agent may device a plan for a goal to be achieved. At some intermediate step in the plan, the next step may no longer be applicable due to (unforeseen) changes in the environment. GOAL tries to avoid the complexity of rebuilding "broken plans" by advocating a reactionary model. We should consider how the agent can select appropriate actions based on the current state of affairs. However, note that it is still possible for programmers to represent plans via the cognitive state of agents using Prolog, but it is not facilitated explicitly by the language. The reactionary approach is not flawless either: it can be difficult for programmers to come up with logical rules that produce the desired behaviour, but by overcoming this challenge, we often have a more flexible agent.

\section{Strategy}\label{strategy}

In our current system, we have a universal agent type. By a universal agent type we mean that all agents share the same logic. While it is possible to have different kinds of agents in GOAL, i.e. via modules, it is not something we currently utilize. Some advantages of a universal agent type are that it is faster to implement, every agent is seamlessly capable of everything and we need not worry about when to switch the agent's type. The main disadvantages are that the code base becomes convoluted as development progresses due to growing array of logical rules for selecting the appropriate action. 

During action selection, the agents apply heuristic measures to determine movement directions. We describe the different variants of heuristic functions in Section \ref{agentmovement}. It should also be noted that we currently do not perform any clear actions. We did not manage to implement use of the action in a meaningful way for the contest.

The following priority list describes the decision-making process of our agents (with some simplifications) where the first applicable rule determines the action to be selected in a given state:
\begin{itemize}
    \item If the agent is assigned to a task:
    \begin{itemize}
        \item Detach any attached blocks not needed for the task. The agent will only detach blocks if it considers it non-obstructive to future movement. If not, it will continue to move (using the detach movement heuristic) until it considers it safe to detach.
        \item Rotate the block into the position dictated by the task plan. If rotation is blocked, move until rotation is possible (using the exploration movement heuristic).
        \item If the agent observes part of the pattern to be handed in, or if the agent is assigned to submit the task (the submitting agent), and is on a goal, wait for other agents (by performing the skip action).
        \item If the agent observes the entire pattern, connect with other blocks/agents as described by the task plan.
        \item If all blocks in the pattern are connected, the assigned agent submits the task.
        \item If the agent finds the submitting agent (waiting in a goal area), move until the attachment(s) form the (partial) pattern (using the task pattern movement heuristic).
        \item If the agent is the submitting agent, move towards a goal area (using the go to movement heuristics).
        \item Else, move towards the position of the submitting agent (using the go to movement heuristics).
        \item If a goal area is known, move towards it (using the go to movement heuristics).
        \item Move into the most promising direction (based on the exploration movement heuristic).
    \end{itemize}
    \bigskip
    \item If the agent is not assigned to a task:
    \begin{itemize}
        \item If a block or dispenser is in vision, and the agent does not have four blocks:
        \begin{itemize}
            \item Rotate such that a free attachment spot is facing the direction of the block/dispenser. If rotation is blocked, move (using the exploration movement heuristic).
            \item If the agent is next to a block, attach it to the agent.
            \item If the agent is next to a dispenser, request a block.
            \item If not next to the block/dispenser, move towards it (using the go to movement heuristics).
        \end{itemize}
        \item Move around on the map (based on the safe exploration heuristic).
    \end{itemize}
    \bigskip
    \item Perform skip action.
\end{itemize}

\section{Agent Knowledge}\label{agentknowledge}

In this section, we cover the design of the knowledge stored in the agents' mental states. Generally speaking, the agents store and maintain knowledge about the map that is assumed be invariable (or alternatively: always perceivable). We consider invariable knowledge to be: the positions of goal cells, attached blocks, the agent's current position (relative to its starting position), positions visited by the agent, and the position of encountered agents from the team. Some of these involve communication between agents. The positions of blocks, dispensers and obstacles are only stored in the agent as long as they are within vision. The communication of our agents is described in Section \ref{agentcomm}. 

The agent does not perceive a global view of the map via the environment, nor its own position on the map. Furthermore, random events and actions of other agents can change the structure of the map over time. Due to this complexity, we do not attempt to build up an internal representation of the map. Unfortunately, this comes at the cost of efficient and meaningful movement on the map.

\subsection{The Current Position}
By keeping track of performed move actions, and checking for a potential failed action, the agent maintains information about its own current position. With the starting position of the agent as the center of origin, we maintain two values that represent the agent's position in a two-dimensional space. Moving in a direction, either north, east, south or west; results in incrementing or decrementing one of these values.



\subsection{Visited Positions}
The bookkeeping of visited positions is essential to avoiding that the agent repeatedly gets stuck, or does not make progress. When an agent performs a move action, in the next step the following information about the visited position is stored: the relative position of the agent, the current step in the simulation, and a flag for if the position is a goal cell. The position is stored relative to the initial position of each agent. That is, each agent is initially at $(x, y) = (0, 0)$. The relative position of each agent is updated based on successful move actions.

As the simulation progresses, the database of visited positions gains additional entries. Due to the way we utilize this, we are only interested in visited positions with respect to a specific subtasks. For instance, if the agent is trying to find a goal cell, it is not relevant which positions the agent visited in an attempt to find blocks. Therefore, we define a number of events that trigger a clearing of the agent's knowledge about visited positions:
\begin{itemize}
    \item The agent has completed a subtask: Attached, detached or requested a block.
    \item We submitted a task. In our system, once a task is submitted, most agents will work on different matters.
\end{itemize}

The visited nodes are useful for steering the agent away from repeating the same movement patterns when they do not make progress. The idea is that visited positions are only relevant locally -- at later points in time it may be relevant to visit those positions again. Essentially, this means that the visited positions are only remembered for the duration of smaller subtasks. Intuitively, it seems non-optimal to remove knowledge that could help steer the agent away from dead ends that it has found earlier. However, the ever-changing structure of the map quickly invalidates this knowledge anyway. 

\subsection{Positions of Goal Cells}
The positions of goal cells are assumed to be invariable throughout the simulation. Due to this assumption, once the agents store knowledge about the positions of goal cells, they are never removed.

The position of goal cells are stored relative to the position of the agents and are thus updated following successful move actions.

The agents learn about positions of goal cells either through perceiving them within their own vision or via communication with other agents.

\subsection{Blocks, Dispensers and Obstacles}
The information about blocks, dispensers and obstacles are only perceived when the agent is within vision. clear events may remove blocks from the map, or they may be moved by other agents. Due to this, the positions of blocks is not maintained when outside of the agent's vision. 

Obstacle positions could potentially be maintained by perceiving clear events and remove information about affected obstacles. The position of obstacles currently plays no part in any sort of route finding algorithm and we do not maintain this knowledge when outside of the agent's vision. 

Dispensers are different from blocks and obstacles as their positions do not change during the simulation. In the current implementation, agents always go towards an available dispenser, if they do not have a block on each side, and if they are not trying to solve a task. Our agents will always prefer to go to the nearest known position of a block or dispenser. Thus to avoid the agent always going back to the same dispenser, we currently do not keep information of dispenser positions outside of the agent's vision. 

Neither the position of blocks, dispenser or obstacles are shared between agents via communication. Since we do not keep and maintain their positions, it does not make sense to share the information between agents -- it should only be part of an agent's mental state when within vision.

\subsection{Attached Blocks}
Each agent keeps track of its own attached blocks with coordinates relative to its own position. The environment makes available any attached blocks in vision, but it is not immediately visible which agents the blocks are attached to. To make sure that the agent only keeps stored knowledge about the blocks attached to itself, we check for successful attach actions to insert the knowledge of a block being attached. Successful rotations update the stored coordinates accordingly. We always make sure that any knowledge about attached blocks is also perceivable in the environment -- if not, the knowledge is removed. This is due to the fact that submitting tasks and clear events may invalidate the knowledge.

We will also briefly mention that current attached blocks are communicated between agents. This is used for devising plans to submit tasks. The details are covered in Sections \ref{agentcomm-attached} and \ref{taskplanning}.

\section{Agent Communication and Shared Knowledge}\label{agentcomm}
Sharing knowledge between agents by means of communication is essential for efficiently exploiting the multiple agents available. The environment presents a number of challenges in enabling effective agent communication. Also, the volatility of the scenario map does not suggest an easy way of building up a shared representation. Our current implementation could utilize more shared knowledge and communication, and it is something we hope to improve in the future.

Specifically for agent programming using GOAL, communication between agents are part of the core loop. One important aspect is that any messages sent in one step will only be available for processing by the receiving agent in the following step. This requires some deliberate implementation to make sure that the information received is up to date -- in our implementation this is extremely relevant as we often share information that is relative to the current position of agents.

\subsection{Encountering other Agents}
The environment only gives information to agents about the position of other agents when within their vision. The agent is able to perceive which team an encountered agent belongs to, but no further identification is provided (i.e. the name of the encountered agent). To be able to identify which pair of agents that have encountered each other we apply the following: when two of our agents meet, they exchange information about what other objects they are able to identify within their vision. Only if they agree on everything in their shared vision, they acknowledge that they did in fact encounter each other. A check is performed to prevent two agents from mistakenly concluding that they encountered each other. We do this by checking that the given pair of agents agree on objects in their shared vision. Our initial implementation was solely based on the agents' relative position to each other, with no additional conditions, which yielded occasional false positives. 

\subsection{Goal Cells and Agent Positions}
When two agents agree that they encountered each other, they exchange information about the positions of goal cells and other agents from the team. Each agent adds the shared information to its belief base relative to its belief about its own current position. Currently, the agents do not continue to share newly information after encountering other agents although this is possible. 

When an agent successfully moves in a given direction, it informs other agents about which directed it moved in. This information is used by each agent to maintain the knowledge about positions of other agents.

\subsection{Attached Blocks}\label{agentcomm-attached} 
We assign one of our agents as the \textit{planning agent}. At each step of the simulation, each agent, that is not currently assigned to solve a task, sends a message to the planning agent containing a list of its currently attached blocks. The planning agent uses the received messages to (possibly) assign a task to a subset of the agents that sent messages. The details of the task planning assignment are covered in Section \ref{taskplanning}.

\section{Agent Movement}\label{agentmovement}
The \textit{Agents Assemble} scenario provides only partial vision of the map to agents, limited to a small area around each agent. Combining the knowledge of agents over time will provide more and more knowledge of the map. However, random clear events happen over time around the map that remove and randomly add new obstacles on part of the map. Other agents also have the ability to remove obstacles. As such, a usual route finding algorithm requires substantial alterations to be usable for the scenario. Due to the volatility of the map, such a route finding algorithm will necessarily have to support re-planning when the planned route is invalidated.

The above mentioned challenges for a route finding algorithm means that we have opted for a more naive implementation of agent movement. The overall strategy is to evaluate each of the (up to four) possible directions: north, east, south and west; and then select the direction which has the optimal heuristic value. When multiple directions share the same optimal value, a direction is selected pseudo-randomly (the current simulation step is used as seed). Our agent movement algorithm has five different variations: 
\begin{itemize}
    \item \textbf{Exploration} favors directions towards positions the agent has not visited recently.
    \item \textbf{Safe exploration} is similar to the above, but further favors directions that increase the distance to goal areas and other agents.
    \item \textbf{Go to} favors directions towards a given relative position and penalizes movement to recently visited positions on the map.
    \item \textbf{Task pattern} favors directions that realize a given task pattern and penalizes movement to recently visited positions on the map.
    \item \textbf{Detach} favors directions away from obstacles.
\end{itemize}

The choice of movement algorithm depends on the current strategy of the agent.

\subsection{Evaluation Functions}
As described above, each variation of the movement algorithm is distinguished by its heuristic function $h$. We will use $h^+$ to denote that a higher value is better and $h^-$ when lower is better. Neither of the mathematical formulations are perfect in any sense, but attempt to give an approximation to the optimal choice. 

\subsubsection{Exploration} 
$$
h^+_{exp}(d) = \sum_{visited}
\left\{ \begin{array}{lr}
\frac{|\Delta x(d)|+|\Delta y(d)|}{\Delta S^2} & \text{for } |\Delta x(d)|+|\Delta y(d)| \leq 30 \text{ and } \Delta S > 0 \\
0 & \text{else}
\end{array} \right.
$$
where $\Delta x(d)$ and $\Delta y(d)$ are the differences in x and y coordinates between the visited position and the agent's position after performing move in direction $d$. $\Delta S$ is the number of steps since the position was visited.

\subsubsection{Safe Exploration}
$$
h^+_{s-exp}(d) = h^+_{exp}(d) + \sum_{(x_t, y_t) \in P} c(x_t, y_t) * ( | x_t|+|y_t|)
$$
where $P$ is a set of coordinates of goal cells and nearby agents in the team. $x_t$ and $y_t$ are coordinates relative to the agent's current position. $c(x_t, y_t)$ is a constant factor used for heavily favoring moving away from nearby agents.

\subsubsection{Go To}
$$
h^-_{go-to}(d) = |\Delta x(d)| + |\Delta y(d)| 
+ \text{size}(V_{p_d})
$$
where $\Delta x(d)$ and $\Delta y(d)$ are the differences in x and y coordinates between the goal position and $p_d$ is the agent's position after performing move in direction $d$. $\text{size}(V_{p_d})$ is the number of times position $p_d$ has been visited recently.
\subsubsection{Task Pattern}
$$
h^-_{pat}(d) =  \text{size}(V_{p_d}) +  \sum_{(x, y, t) \in (pat/att)} min \left\{ |\Delta x| + |\Delta y| \;\mid\; \Psi(d, x, y, t, \Delta x, \Delta y) \right\}
$$
where $\text{size}(V_{p_d})$ is the number of times $p_d$ has been visited recently ($p_d$ is the agent's position after performing move in direction $d$). The set $pat/att$ contains the relative position and type of every block in the pattern excluding the blocks the agent itself is providing (has attached). The predicate $\Psi(d, x, y, t, \Delta x, \Delta y)$ gives the difference in x and y coordinates to every observed non-attached block of type $t$ assuming a move in direction $d$.
\subsubsection{Detach}
$$
h^+_{det}(d) = h^+_{exp}(d) + \sum_{(x,y) \in \text{obstacles}} |\Delta x(d)| + |\Delta y(d)|
$$
where $\text{obstacles}$ is the set of the positions of observable obstacles. $|\Delta x(d)|$ and $|\Delta y(d)|$ are the relative differences in x and y coordinates between the agent and the obstacle following a move in direction $d$.

\section{Solving Tasks}\label{solvingtasks}
This section describes how the agents solve tasks. We consider solving a task to consist of four parts: Collecting blocks, planning tasks to complete based on the collected blocks, executing task plans (assembling the pattern) and finally submitting the pattern.

\subsection{Collecting Blocks}
One core aspect of our strategy is to collect blocks before committing to any of the available tasks, and to only commit to tasks for which we already have the blocks to solve.

If an agent, that is not assigned to a task, and does not hold a block on each side, encounters a block or dispenser, it will generally try to go towards it. It will only ignore the possibility to collect the block(s) in case another from the same team is adjacent to it. This is to avoid race conditions for the same resource and improve efficiency. Dealing with this issue via communication would likely be a better approach, however.

In case the agent sees a dispenser or block, that is not occupied by another agent from the same team, the agent will rotate if necessary to ensure that a free attachment spot is available in the direction of the block or dispenser. Attaching blocks takes priority over requesting blocks from dispensers. The agent will repeatedly attach blocks on each of the four spots until they are all used. 

In some cases the position of the block or dispenser may not allow the agent to attach from that angle due to its current attachments. This is not currently checked and avoided. In such a case, the agent is likely to enter a state of not making progress -- unless it is assigned to a task, or if it moves outside of vision of the block due to being penalized for going to similar positions repeatedly.

\subsection{Task Planning}\label{taskplanning}
If the agents combined have the attachments needed to solve a task, the planning agent computes a task plan for it. The task with the lowest reward (and thus, likely the easiest to complete) is selected. We select the easiest task based on the observed performance of the agents, and we only ever try to complete one task at a time. The logic for task planning supports computing multiple non-overlapping task plans, but we observed that the agents would be likely to obstruct each other. We only ever commit to a task that has some amount of steps available to complete the task before the deadline. For the contest, we set this to a minimum of 50 steps to complete the task. Later experimentation has yielded better results with a higher number. Due to the abundance of available tasks, the improved results seem logical. So far we have not conducted tests of the distribution of task completion times for our agents.

The task plan specifies how each agent provides part of the pattern including how it should be rotated. The expected completed pattern is computed for each agent relative to its own position in the aligned pattern. This is used to ensure that the agents align their attachments correctly. The task plan also specifies how agents should connect to each other once the pattern alignment step is complete. For each task, one of the selected agents is assigned to submit the task.

The task plans are rigid in the sense that two agents that provide the same block type cannot swap their respective sub-patterns. While there potentially could be some benefit in supporting this behaviour, we do not consider it worth the effort to implement when considering other potential improvements to the agents.

\subsection{Executing Task Plans}
When a task plan is sent to the agents, the agents not involved will keep on moving around the map, but their heuristics for movement will penalize moving close to other agents. This is done in an attempt to avoid obstructing agents that are working on completing a task. Each of the involved agents will immediately detach any of the blocks that are no longer needed and rotate the remaining attachments to align with their part of the pattern for the task. The agent responsible for submitting the task will move towards the nearest goal cell if its position is known (or scout for a goal cell if not). Once positioned at a goal cell, the agent will wait for the other agents to show up. If the other agents assigned to the task know the position of the agent responsible to submitting the task, they will move towards that agent. If not, they will methodically visit each of the known goal areas. If this fails, they will scout around the map. In most practical cases, the agents eventually learn of the position of the agent responsible for submitting the task.

It should be noted that problems due to assigned agents that become stuck is not currently resolved. However, if one of the assigned agents gets disabled, due to a clear event, the task plan is deleted. There is implemented some logic to check if an agent is stuck, in which case the task plan is to be deleted, but it is not currently functional. As previously mentioned, we also do not utilize the clear action in any way currently, which could solve potential pathing problems. At any time, if the task deadline is exceeded, the task plan is deleted.

In case an agent assigned to the task finds its way to the agent responsible for submitting the task, waiting at a goal area, the agent will start to align itself to complete the pattern. This is achieved by the task pattern heuristic that favors directions that minimize the expected deviance from the pattern to submit. It should be noted that the agent responsible for submitting the task will try to position itself such that the other agents have room to align themselves to complete the pattern.

Once the pattern is complete, the agents connect their attachments and the task is submitted.

\section{Evaluation of Matches}\label{evaluation}
With a total of four participants, we played three matches against each of the three opponents. In the following, we evaluate the performance of our agents in each matchup. See Figures \ref{plot-trg-score-1}-\ref{plot-lfc-clear-3} for key statistics over the 500 steps of each match.

\subsection*{GOAL-DTU vs. TRG}
In two of the simulations, we manage to complete a single task early on (Figures \ref{plot-trg-tasks-1}, \ref{plot-trg-tasks-2}, \ref{plot-trg-tasks-3}). Team TRG completes a single task in the first simulation, but are unable to do so in the other simulations. TRG has a strategy where some of their agents defend goal areas by attempting to perform clear actions on our agents trying to complete tasks. Since part of our task execution plan is to assemble the pattern in the goal area, the strategy of team TRG denies a fair number of submits from our agents.

We experience problems with many agent getting stuck in every simulation. Around halfway through, we can usually observe that half of our agents are now stuck. This is especially detrimental if one of those agents is assigned to complete a task.

It seems that our greedy approach for collecting blocks, which causes the map to become even more convoluted, also causes serious issues for the agents of team TRG.

In the third simulation, the number of blocks we manage to collect stagnates (Figure \ref{plot-trg-blocks-3}). This could be correlated with the number of clear events that is significantly higher (Figure \ref{plot-trg-clear-3}). While this presumably has no direct impact our score, it could be an indication towards our agents' ability to move around on the map.

See Figures \ref{plot-trg-score-1}-\ref{plot-trg-clear-3} for all of the collected statistics in matches vs. TRG.

\subsection*{GOAL-DTU vs. FIT BUT}
For this matchup we experienced issues with the agents. In the first simulation, our agents manage to assemble two patterns for tasks, but in one instance they seem to have a wrong pattern, and in the other instance they try to submit outside of a goal area (Figure \ref{plot-fitbut-submit-1}). At this point, we try to restart our agents, but they do not mange to make meaningful progress since our implementation is not robust in case of crashes (Figure \ref{plot-fitbut-blocks-1}).

Also in the second simulation we have to attempt a restart, but to no avail. It seems our agents obstruct the map so severely that team FIT BUT has issues (Figure \ref{plot-fitbut-tasks-2}). Before our agents crash, they are relatively close to assembling a pattern.

The story repeats itself in the third simulation, although team FIT BUT successfully complete multiple tasks (Figure \ref{plot-fitbut-tasks-3}).

While we do not expect us to have been able to beat the agents of team FIT BUT, we would likely have been able to complete a few tasks if the agents did not crash.

See Figures \ref{plot-fitbut-score-1}-\ref{plot-fitbut-clear-3} for all of the collected statistics in matches vs. FIT BUT.

\subsection*{LFC vs. GOAL-DTU}
We managed to complete tasks in each of the three simulations (Figures \ref{plot-lfc-taks-1}, \ref{plot-lfc-taks-2}, \ref{plot-lfc-taks-3}). Yet, as for other simulations, as the simulation progresses, the agents' ability to move around the map degrades (Figures \ref{plot-lfc-blocks-1}, \ref{plot-lfc-blocks-2}, \ref{plot-lfc-blocks-3}). In comparison, it seems that the agents of team LFC make steady progress throughout the simulation (Figures \ref{plot-lfc-taks-1}, \ref{plot-lfc-taks-2}, \ref{plot-lfc-taks-3}). Another note about the agents of team LFC is there seems to be a correlation between the tasks they submit and the number of blocks they collect which suggest a different strategy (Figures \ref{plot-lfc-blocks-1}, \ref{plot-lfc-blocks-2}, \ref{plot-lfc-blocks-3}).

We manage to complete more tasks in the second simulation (Figure \ref{plot-lfc-taks-2}). By inspection of the map layout, there are two goal areas in the middle of map, not close to any obstacles. This is a lucky coincidence for our agents, as they often experience more problems when close to obstacles (maneuverability in confined spaces is more likely to degrade over time).

See Figures \ref{plot-lfc-score-1}-\ref{plot-lfc-clear-3} for all of the collected statistics in matches vs. LFC.

\section{Discussion}\label{discussion}
While our implementation achieved satisfying results, it can still be improved on several fronts. During the contest, we learned that the performance of the agents could be improved by tinkering with parameters, while other issues were due to technical difficulties. 

\subsection{Changes since the Contest}
After analysis of the replays of our matches in the contest, we realized two possible improvements to the implementation. The first improvement is concerned with the assumed time our agents need to complete a task. Through experimentation, we have learned that increasing the minimum amount of steps needed to complete a task improved the performance of the agents significantly. The value used for the matches in the contest often lead to agents missing task deadlines resulting in a lower score, when considering that a task with a later deadline could have been chosen instead.

Another issue occurs when agents detach blocks (for getting rid of blocks not needed for completing the assigned task). The agents will try to detach the unneeded blocks away from goal areas and obstacles. This is done to avoid that the agent potentially obstructs itself and other agents. By increasing the minimum distance there should be to goal cells and obstacles when detaching a block, we observed improved performance.

\subsection{Technical Issues during the Contest}
During the contest, we experienced a number of issues during the simulations that we had not encountered before. Unfortunately, for some of the matches this made our agents break down completely, practically leaving us with no way to continue. Since we did not experience this before, our agents are not very robust in the sense that they are not well-suited for restarts during the simulation in case of crashes. 

One of the issues we experienced occurs when the steps are performed very rapidly, at which point it seems as if the GOAL execution and the server get out of sync. We did not experience this problem earlier since in our testing, there were always agents not performing actions, thus using the full server timeout for each step. Experimentation following the contest shows that the problem is not related to connection issues and will have to be investigated further. We find that two seconds for each step prevents the issue.

\subsection{Known Problems and Bugs}
In our implementation, we have discovered a number of problems over time, and there are still some unresolved bugs to fix.

One problem is related to agents getting stuck. Often in simulations, we will experience that one or more of our agents end up getting stuck. This could be due to a number of random clear events, or potentially the map has disconnected parts. One of the major issues with this is that we have not been able to implement a way for agents to deduce that they are in fact stuck, or that some agents are unable to reach each other. Another problem is that we do not utilize the clear action to help the agents become unstuck.

Lastly, we have experienced problems when agents assigned to a task visit goal areas in search for the agent responsible for submitting the task. Due to an unresolved bug, the agents will not properly scout all the goal areas, but tend to always stay in the same area. This obviously means that we will never be able to submit unless the problematic agents learn about the position of the agent responsible for submitting the task.

\subsection{Improvements}
There are a number of directions to take in terms of improving the implementation. We will consider some of our high-level ideas to improve the performance of the system by targeting some of our weaknesses.

Our agents are universal in the sense that every agent is based on exactly the same logic rules. One way to improve the performance, could be to assign different roles to agents, or to assign agents into smaller teams that move together. Examples of roles could be agents that explore the map, some that request and collect blocks and some that complete tasks using the collected blocks. 

Another weakness of our agents is poor movement. Since we do not build an internal representation of the map during the simulation, our agents often move in the blind. We expect substantially improved performance if we are able to make the agents better at moving around the map, for example by building up such an internal map representation (which could then be shared among agents). The initial reason not to attempt building up a map representation, is the complexity of the map being dynamic. 

Another problem arises from our greedy approach to collecting blocks. Since each agent always tries to collect one block on each side, movement around the map becomes much harder afterwards. Furthermore, we do not consider the possibility that an agent may be able to move past narrow corridors by rotating its attachments, or potentially even moving the blocks past corridors a few at a time.

The last obvious improvement is to implement some logic to perform clear actions. Clear events are likely to make moving around the map very difficult over time, and clearing up obstructions is exactly one of the purposes of the clear action. It can be used to reconnect parts of the map that has been disconnected completely, and to save valuable time by creating shortcuts through obstacles.

Lastly, there are number of technical improvements that we would like to implement. Since it was impossible to monitor the agents live during the simulations of the contest, it would have been helpful to have better output (in the console) about the behaviour and progress of the agents. Furthermore, we would like to increase the robustness of the agents in case of crashes such that they can be restarted and still make progress.

\section{Conclusion}\label{conclusion}
We have provided an overview of the multi-agent system that the GOAL-DTU team developed for the Multi-Agent Programming Contest 2019. We have explained our choice of the GOAL programming language; we have also described the main strategy of our agents and how they execute that strategy. This year was the first iteration using the \textit{Agents Assemble} scenario, and we have developed our implementation from scratch using GOAL. Our implementation features a universal agent type in which each agent is based on the same set of logical rules.

The strengths of our system are the flexible nature of our agents. Our agents always react to the current state of affairs and do not rely heavily on predefined plans to reach their goal of completing tasks. The weaknesses are primarily the agents' poor movement around the map and rigidity in the way tasks are assigned and submitted where stuck agents have a severe negative impact on the performance of the system. 

Finally, we have described how to improve the system by coming up with ideas that target its weaknesses. Some of these potential improvements are related to minor issues and bug fixes while other potential improvements require designing and refactoring large parts of the system.

In conclusion, we are satisfied with the performance of our system, ending at a 3rd place in the final rankings, when considering that we have built the system from scratch. We consider our current implementation a good platform to built on for future iterations of the \textit{Agents Assemble} scenario.

\

\vfill

\begin{center}
Further details about the previous DTU teams are available here:
\\[3ex]
\url{https://people.compute.dtu.dk/jovi/MAS/}
\end{center} 

\newpage

\begin{figure}[H]
    \centering
    \begin{minipage}{0.45\textwidth}
        \centering
        
        \resizebox{\textwidth}{!}{
        \begin{tikzpicture}
        \begin{axis}[
        xlabel=Steps,
        ylabel=Score,
        xmin=0, xmax=500,
        ymin=0, ymax=100,
        xtick={0,100,...,500},
        ytick={0,10,...,100},
        ymajorgrids=true,
        xmajorgrids=true,
        grid style=dashed,
        ]
        \addplot+[line width=2pt,mark=none] table [x=Step, y=DTU-Score]{plotdata/11.dat};
        \addlegendentry{GOAL-DTU}
        \addplot+[line width=2pt,mark=none,opacity=0.55] table [x=Step, y=VS-Score]{plotdata/11.dat};
        \addlegendentry{TRG}
        \end{axis}
        \end{tikzpicture}
        }
        
        \caption{Score: GOAL-DTU vs. TRG (1)}
        \label{plot-trg-score-1}
    \end{minipage}\hfill
    \begin{minipage}{0.45\textwidth}
        \centering

        \resizebox{\textwidth}{!}{
        \begin{tikzpicture}
        \begin{axis}[
        xlabel=Steps,
        ylabel=Score,
        xmin=0, xmax=500,
        ymin=0, ymax=100,
        xtick={0,100,...,500},
        ytick={0,10,...,100},
        ymajorgrids=true,
        xmajorgrids=true,
        grid style=dashed,
        ]
        \addplot+[line width=2pt,mark=none] table [x=Step, y=DTU-Score]{plotdata/12.dat};
        \addlegendentry{GOAL-DTU}
        \addplot+[line width=2pt,mark=none,opacity=0.55] table [x=Step, y=VS-Score]{plotdata/12.dat};
        \addlegendentry{TRG}
        \end{axis}
        \end{tikzpicture}
        }

        \caption{Score: GOAL-DTU vs. TRG (2)}
        \label{plot-trg-score-2}
    \end{minipage}
    
    \vspace{5.5em}
    
    \begin{minipage}{0.45\textwidth}
        \centering
        
        \resizebox{\textwidth}{!}{
        \begin{tikzpicture}
        \begin{axis}[
        xlabel=Steps,
        ylabel=Score,
        xmin=0, xmax=500,
        ymin=0, ymax=100,
        xtick={0,100,...,500},
        ytick={0,10,...,100},
        ymajorgrids=true,
        xmajorgrids=true,
        grid style=dashed,
        ]
        \addplot+[line width=2pt,mark=none] table [x=Step, y=DTU-Score]{plotdata/13.dat};
        \addlegendentry{GOAL-DTU}
        \addplot+[line width=2pt,mark=none,opacity=0.55] table [x=Step, y=VS-Score]{plotdata/13.dat};
        \addlegendentry{TRG}
        \end{axis}
        \end{tikzpicture}
        }
        
        \caption{Score: GOAL-DTU vs. TRG (3)}
        \label{plot-trg-score-3}
    \end{minipage}\hfill
    \begin{minipage}{0.45\textwidth}
        \centering

        \resizebox{\textwidth}{!}{
        
        \begin{tikzpicture}
        \begin{axis}[
        xlabel=Steps,
        ylabel=Total blocks attached,
        xmin=0, xmax=500,
        ymin=0, ymax=150,
        xtick={0,100,...,500},
        ytick={0,25,...,150},
        ymajorgrids=true,
        xmajorgrids=true,
        grid style=dashed,
        ]
        \addplot+[line width=2pt,mark=none] table [x=Step, y=DTU-AttachedBlocks]{plotdata/11.dat};
        \addlegendentry{GOAL-DTU}
        \addplot+[line width=2pt,mark=none,opacity=0.55] table [x=Step, y=VS-AttachedBlocks]{plotdata/11.dat};
        \addlegendentry{TRG}
        \end{axis}
        \end{tikzpicture}
        }

        \caption{Blocks: GOAL-DTU vs. TRG (1)}
        \label{plot-trg-blocks-1}
    \end{minipage}
    
    \vspace{5.5em}

    \begin{minipage}{0.45\textwidth}
        \centering
        
        \resizebox{\textwidth}{!}{
        \begin{tikzpicture}
        \begin{axis}[
        xlabel=Steps,
        ylabel=Total blocks attached,
        xmin=0, xmax=500,
        ymin=0, ymax=150,
        xtick={0,100,...,500},
        ytick={0,25,...,150},
        ymajorgrids=true,
        xmajorgrids=true,
        grid style=dashed,
        ]
        \addplot+[line width=2pt,mark=none] table [x=Step, y=DTU-AttachedBlocks]{plotdata/12.dat};
        \addlegendentry{GOAL-DTU}
        \addplot+[line width=2pt,mark=none,opacity=0.55] table [x=Step, y=VS-AttachedBlocks]{plotdata/12.dat};
        \addlegendentry{TRG}
        \end{axis}
        \end{tikzpicture}
        }
        
        \caption{Blocks: GOAL-DTU vs. TRG (2)}
        \label{plot-trg-blocks-2}
    \end{minipage}\hfill
    \begin{minipage}{0.45\textwidth}
        \centering

        \resizebox{\textwidth}{!}{
        \begin{tikzpicture}
        \begin{axis}[
        xlabel=Steps,
        ylabel=Total blocks attached,
        xmin=0, xmax=500,
        ymin=0, ymax=150,
        xtick={0,100,...,500},
        ytick={0,25,...,150},
        ymajorgrids=true,
        xmajorgrids=true,
        grid style=dashed,
        ]
        \addplot+[line width=2pt,mark=none] table [x=Step, y=DTU-AttachedBlocks]{plotdata/13.dat};
        \addlegendentry{GOAL-DTU}
        \addplot+[line width=2pt,mark=none,opacity=0.55] table [x=Step, y=VS-AttachedBlocks]{plotdata/13.dat};
        \addlegendentry{TRG}
        \end{axis}
        \end{tikzpicture}
        }

        \caption{Blocks: GOAL-DTU vs. TRG (3)}
        \label{plot-trg-blocks-3}
    \end{minipage}
\end{figure}

\begin{figure}[H]
    \centering
    \begin{minipage}{0.45\textwidth}
        \centering
        
        \resizebox{\textwidth}{!}{
        \begin{tikzpicture}
        \begin{axis}[
        xlabel=Steps,
        ylabel=Total submits attempted,
        xmin=0, xmax=500,
        ymin=0, ymax=5,
        xtick={0,100,...,500},
        ytick={0,1,...,5},
        ymajorgrids=true,
        xmajorgrids=true,
        grid style=dashed,
        ]
        \addplot+[line width=2pt,mark=none] table [x=Step, y=DTU-SubmitAttempts]{plotdata/11.dat};
        \addlegendentry{GOAL-DTU}
        \addplot+[line width=2pt,mark=none,opacity=0.55] table [x=Step, y=VS-SubmitAttempts]{plotdata/11.dat};
        \addlegendentry{TRG}
        \end{axis}
        \end{tikzpicture}
        }
        
        \caption{Submit: GOAL-DTU vs. TRG (1)}
        \label{plot-trg-submit-1}
    \end{minipage}\hfill
    \begin{minipage}{0.45\textwidth}
        \centering

        \resizebox{\textwidth}{!}{
        \begin{tikzpicture}
        \begin{axis}[
        xlabel=Steps,
        ylabel=Total submits attempted,
        xmin=0, xmax=500,
        ymin=0, ymax=5,
        xtick={0,100,...,500},
        ytick={0,1,...,5},
        ymajorgrids=true,
        xmajorgrids=true,
        grid style=dashed,
        ]
        \addplot+[line width=2pt,mark=none] table [x=Step, y=DTU-SubmitAttempts]{plotdata/12.dat};
        \addlegendentry{GOAL-DTU}
        \addplot+[line width=2pt,mark=none,opacity=0.55] table [x=Step, y=VS-SubmitAttempts]{plotdata/12.dat};
        \addlegendentry{TRG}
        \end{axis}
        \end{tikzpicture}
        }

        \caption{Submit: GOAL-DTU vs. TRG (2)}
        \label{plot-trg-submit-2}
    \end{minipage}
    
    \vspace{5.5em}
    
    \begin{minipage}{0.45\textwidth}
        \centering
        
        \resizebox{\textwidth}{!}{
        \begin{tikzpicture}
        \begin{axis}[
        xlabel=Steps,
        ylabel=Total submits attempted,
        xmin=0, xmax=500,
        ymin=0, ymax=5,
        xtick={0,100,...,500},
        ytick={0,1,...,5},
        ymajorgrids=true,
        xmajorgrids=true,
        grid style=dashed,
        ]
        \addplot+[line width=2pt,mark=none] table [x=Step, y=DTU-SubmitAttempts]{plotdata/13.dat};
        \addlegendentry{GOAL-DTU}
        \addplot+[line width=2pt,mark=none,opacity=0.55] table [x=Step, y=VS-SubmitAttempts]{plotdata/13.dat};
        \addlegendentry{TRG}
        \end{axis}
        \end{tikzpicture}
        }
        
        \caption{Submit: GOAL-DTU vs. TRG (3)}
        \label{plot-trg-submit-3}
    \end{minipage}\hfill
    \begin{minipage}{0.45\textwidth}
        \centering

        \resizebox{\textwidth}{!}{
        
        \begin{tikzpicture}
        \begin{axis}[
        xlabel=Steps,
        ylabel=Total tasks completed,
        xmin=0, xmax=500,
        ymin=0, ymax=5,
        xtick={0,100,...,500},
        ytick={0,1,...,5},
        ymajorgrids=true,
        xmajorgrids=true,
        grid style=dashed,
        ]
        \addplot+[line width=2pt,mark=none] table [x=Step, y=DTU-CompletedTasks]{plotdata/11.dat};
        \addlegendentry{GOAL-DTU}
        \addplot+[line width=2pt,mark=none,opacity=0.55] table [x=Step, y=VS-CompletedTasks]{plotdata/11.dat};
        \addlegendentry{TRG}
        \end{axis}
        \end{tikzpicture}
        }

        \caption{Tasks: GOAL-DTU vs. TRG (1)}
        \label{plot-trg-tasks-1}
    \end{minipage}
    
    \vspace{5.5em}

    \begin{minipage}{0.45\textwidth}
        \centering
        
        \resizebox{\textwidth}{!}{
        \begin{tikzpicture}
        \begin{axis}[
        xlabel=Steps,
        ylabel=Total tasks completed,
        xmin=0, xmax=500,
        ymin=0, ymax=5,
        xtick={0,100,...,500},
        ytick={0,1,...,5},
        ymajorgrids=true,
        xmajorgrids=true,
        grid style=dashed,
        ]
        \addplot+[line width=2pt,mark=none] table [x=Step, y=DTU-CompletedTasks]{plotdata/12.dat};
        \addlegendentry{GOAL-DTU}
        \addplot+[line width=2pt,mark=none,opacity=0.55] table [x=Step, y=VS-CompletedTasks]{plotdata/12.dat};
        \addlegendentry{TRG}
        \end{axis}
        \end{tikzpicture}
        }
        
        \caption{Tasks: GOAL-DTU vs. TRG (2)}
        \label{plot-trg-tasks-2}
    \end{minipage}\hfill
    \begin{minipage}{0.45\textwidth}
        \centering

        \resizebox{\textwidth}{!}{
        \begin{tikzpicture}
        \begin{axis}[
        xlabel=Steps,
        ylabel=Total tasks completed,
        xmin=0, xmax=500,
        ymin=0, ymax=5,
        xtick={0,100,...,500},
        ytick={0,1,...,5},
        ymajorgrids=true,
        xmajorgrids=true,
        grid style=dashed,
        ]
        \addplot+[line width=2pt,mark=none] table [x=Step, y=DTU-CompletedTasks]{plotdata/13.dat};
        \addlegendentry{GOAL-DTU}
        \addplot+[line width=2pt,mark=none,opacity=0.55] table [x=Step, y=VS-CompletedTasks]{plotdata/13.dat};
        \addlegendentry{TRG}
        \end{axis}
        \end{tikzpicture}
        }

        \caption{Tasks: GOAL-DTU vs. TRG (3)}
        \label{plot-trg-tasks-3}
    \end{minipage}
\end{figure}

\begin{figure}[H]
    \centering
    \begin{minipage}{0.45\textwidth}
        \centering
        
        \resizebox{\textwidth}{!}{
        \begin{tikzpicture}
        \begin{axis}[
        xlabel=Steps,
        ylabel=Clear events,
        xmin=0, xmax=500,
        ymin=0, ymax=50,
        xtick={0,100,...,500},
        ytick={0,10,...,50},
        ymajorgrids=true,
        xmajorgrids=true,
        grid style=dashed,
        ]
        \addplot+[line width=2pt,mark=none] table [x=Step, y=ClearEvents]{plotdata/11.dat};
        \end{axis}
        \end{tikzpicture}
        }
        
        \caption{Clear: GOAL-DTU vs. TRG (1)}
        \label{plot-trg-clear-1}
    \end{minipage}\hfill
    \begin{minipage}{0.45\textwidth}
        \centering

        \resizebox{\textwidth}{!}{
        
        \begin{tikzpicture}
        \begin{axis}[
        xlabel=Steps,
        ylabel=Clear events,
        xmin=0, xmax=500,
        ymin=0, ymax=50,
        xtick={0,100,...,500},
        ytick={0,10,...,50},
        ymajorgrids=true,
        xmajorgrids=true,
        grid style=dashed,
        ]
        \addplot+[line width=2pt,mark=none] table [x=Step, y=ClearEvents]{plotdata/12.dat};
        \end{axis}
        \end{tikzpicture}
        }

        \caption{Clear: GOAL-DTU vs. TRG (2)}
        \label{plot-trg-clear-2}
    \end{minipage}
    
    \vspace{5.5em}

    \begin{minipage}{0.45\textwidth}
        \centering
        
        \resizebox{\textwidth}{!}{
        \begin{tikzpicture}
        \begin{axis}[
        xlabel=Steps,
        ylabel=Clear events,
        xmin=0, xmax=500,
        ymin=0, ymax=50,
        xtick={0,100,...,500},
        ytick={0,10,...,50},
        ymajorgrids=true,
        xmajorgrids=true,
        grid style=dashed,
        ]
        \addplot+[line width=2pt,mark=none] table [x=Step, y=ClearEvents]{plotdata/13.dat};
        \end{axis}
        \end{tikzpicture}
        }

        \caption{Clear: GOAL-DTU vs. TRG (3)}
        \label{plot-trg-clear-3}
    \end{minipage}\hfill
    \begin{minipage}{0.45\textwidth}
        \centering

        \resizebox{\textwidth}{!}{
        \begin{tikzpicture}
        \begin{axis}[
        xlabel=Steps,
        ylabel=Score,
        xmin=0, xmax=500,
        ymin=0, ymax=900,
        xtick={0,100,...,500},
        ytick={0,100,...,900},
        ymajorgrids=true,
        xmajorgrids=true,
        grid style=dashed,
        ]
        \addplot+[line width=2pt,mark=none] table [x=Step, y=DTU-Score]{plotdata/21.dat};
        \addlegendentry{GOAL-DTU}
        \addplot+[line width=2pt,mark=none,opacity=0.55] table [x=Step, y=VS-Score]{plotdata/21.dat};
        \addlegendentry{FIT BUT}
        \end{axis}
        \end{tikzpicture}
        }

        \caption{Score: GOAL-DTU vs. FIT BUT (1)}
        \label{plot-fitbut-score-1}
    \end{minipage}
    
    \vspace{5.5em}

    \begin{minipage}{0.45\textwidth}
        \centering

        \resizebox{\textwidth}{!}{
        \begin{tikzpicture}
        \begin{axis}[
        xlabel=Steps,
        ylabel=Score,
        xmin=0, xmax=500,
        ymin=0, ymax=900,
        xtick={0,100,...,500},
        ytick={0,100,...,900},
        ymajorgrids=true,
        xmajorgrids=true,
        grid style=dashed,
        ]
        \addplot+[line width=2pt,mark=none] table [x=Step, y=DTU-Score]{plotdata/22.dat};
        \addlegendentry{GOAL-DTU}
        \addplot+[line width=2pt,mark=none,opacity=0.55] table [x=Step, y=VS-Score]{plotdata/22.dat};
        \addlegendentry{FIT BUT}
        \end{axis}
        \end{tikzpicture}
        }

        \caption{Score: GOAL-DTU vs. FIT BUT (2)}
        \label{plot-fitbut-score-2}
    \end{minipage}\hfill
    \begin{minipage}{0.45\textwidth}
        \centering

        \resizebox{\textwidth}{!}{
        \begin{tikzpicture}
        \begin{axis}[
        xlabel=Steps,
        ylabel=Score,
        xmin=0, xmax=500,
        ymin=0, ymax=900,
        xtick={0,100,...,500},
        ytick={0,100,...,900},
        ymajorgrids=true,
        xmajorgrids=true,
        grid style=dashed,
        ]
        \addplot+[line width=2pt,mark=none] table [x=Step, y=DTU-Score]{plotdata/23.dat};
        \addlegendentry{GOAL-DTU}
        \addplot+[line width=2pt,mark=none,opacity=0.55] table [x=Step, y=VS-Score]{plotdata/23.dat};
        \addlegendentry{FIT BUT}
        \end{axis}
        \end{tikzpicture}
        }

        \caption{Score: GOAL-DTU vs. FIT BUT (3)}
        \label{plot-fitbut-score-3}
    \end{minipage}
\end{figure}

\begin{figure}[H]
    \centering
    \begin{minipage}{0.45\textwidth}
        \centering

        \resizebox{\textwidth}{!}{
        
        \begin{tikzpicture}
        \begin{axis}[
        xlabel=Steps,
        ylabel=Total blocks attached,
        xmin=0, xmax=500,
        ymin=0, ymax=150,
        xtick={0,100,...,500},
        ytick={0,25,...,150},
        ymajorgrids=true,
        xmajorgrids=true,
        grid style=dashed,
        ]
        \addplot+[line width=2pt,mark=none] table [x=Step, y=DTU-AttachedBlocks]{plotdata/21.dat};
        \addlegendentry{GOAL-DTU}
        \addplot+[line width=2pt,mark=none,opacity=0.55] table [x=Step, y=VS-AttachedBlocks]{plotdata/21.dat};
        \addlegendentry{FIT BUT}
        \end{axis}
        \end{tikzpicture}
        }

        \caption{Blocks: GOAL-DTU vs. FIT BUT (1)}
        \label{plot-fitbut-blocks-1}
    \end{minipage}\hfill
    \begin{minipage}{0.45\textwidth}
        \centering

        \resizebox{\textwidth}{!}{
        
        \begin{tikzpicture}
        \begin{axis}[
        xlabel=Steps,
        ylabel=Total blocks attached,
        xmin=0, xmax=500,
        ymin=0, ymax=150,
        xtick={0,100,...,500},
        ytick={0,25,...,150},
        ymajorgrids=true,
        xmajorgrids=true,
        grid style=dashed,
        ]
        \addplot+[line width=2pt,mark=none] table [x=Step, y=DTU-AttachedBlocks]{plotdata/22.dat};
        \addlegendentry{GOAL-DTU}
        \addplot+[line width=2pt,mark=none,opacity=0.55] table [x=Step, y=VS-AttachedBlocks]{plotdata/22.dat};
        \addlegendentry{FIT BUT}
        \end{axis}
        \end{tikzpicture}
        }

        \caption{Blocks: GOAL-DTU vs. FIT BUT (2)}
        \label{plot-fitbut-blocks-2}
    \end{minipage}
    
    \vspace{5.5em}
    
    \begin{minipage}{0.45\textwidth}
        \centering

        \resizebox{\textwidth}{!}{
        
        \begin{tikzpicture}
        \begin{axis}[
        xlabel=Steps,
        ylabel=Total blocks attached,
        xmin=0, xmax=500,
        ymin=0, ymax=150,
        xtick={0,100,...,500},
        ytick={0,25,...,150},
        ymajorgrids=true,
        xmajorgrids=true,
        grid style=dashed,
        ]
        \addplot+[line width=2pt,mark=none] table [x=Step, y=DTU-AttachedBlocks]{plotdata/23.dat};
        \addlegendentry{GOAL-DTU}
        \addplot+[line width=2pt,mark=none,opacity=0.55] table [x=Step, y=VS-AttachedBlocks]{plotdata/23.dat};
        \addlegendentry{FIT BUT}
        \end{axis}
        \end{tikzpicture}
        }

        \caption{Blocks: GOAL-DTU vs. FIT BUT (3)}
        \label{plot-fitbut-blocks-3}
    \end{minipage}\hfill
    \begin{minipage}{0.45\textwidth}
        \centering
        
        \resizebox{\textwidth}{!}{
        \begin{tikzpicture}
        \begin{axis}[
        xlabel=Steps,
        ylabel=Total submits attempted,
        xmin=0, xmax=500,
        ymin=0, ymax=50,
        xtick={0,100,...,500},
        ytick={0,10,...,50},
        ymajorgrids=true,
        xmajorgrids=true,
        grid style=dashed,
        ]
        \addplot+[line width=2pt,mark=none] table [x=Step, y=DTU-SubmitAttempts]{plotdata/21.dat};
        \addlegendentry{GOAL-DTU}
        \addplot+[line width=2pt,mark=none,opacity=0.55] table [x=Step, y=VS-SubmitAttempts]{plotdata/21.dat};
        \addlegendentry{FIT BUT}
        \end{axis}
        \end{tikzpicture}
        }
        
        \caption{Submit: GOAL-DTU vs. FIT BUT (1)}
        \label{plot-fitbut-submit-1}
    \end{minipage}
    
    \vspace{5.5em}
        
    \begin{minipage}{0.45\textwidth}
        \centering

        \resizebox{\textwidth}{!}{
        \begin{tikzpicture}
        \begin{axis}[
        xlabel=Steps,
        ylabel=Total submits attempted,
        xmin=0, xmax=500,
        ymin=0, ymax=50,
        xtick={0,100,...,500},
        ytick={0,10,...,50},
        ymajorgrids=true,
        xmajorgrids=true,
        grid style=dashed,
        ]
        \addplot+[line width=2pt,mark=none] table [x=Step, y=DTU-SubmitAttempts]{plotdata/22.dat};
        \addlegendentry{GOAL-DTU}
        \addplot+[line width=2pt,mark=none,opacity=0.55] table [x=Step, y=VS-SubmitAttempts]{plotdata/22.dat};
        \addlegendentry{FIT BUT}
        \end{axis}
        \end{tikzpicture}
        }

        \caption{Submit: GOAL-DTU vs. FIT BUT (2)}
        \label{plot-fitbut-submit-2}
    \end{minipage}\hfill
    \begin{minipage}{0.45\textwidth}
        \centering
        
        \resizebox{\textwidth}{!}{
        \begin{tikzpicture}
        \begin{axis}[
        xlabel=Steps,
        ylabel=Total submits attempted,
        xmin=0, xmax=500,
        ymin=0, ymax=50,
        xtick={0,100,...,500},
        ytick={0,10,...,50},
        ymajorgrids=true,
        xmajorgrids=true,
        grid style=dashed,
        ]
        \addplot+[line width=2pt,mark=none] table [x=Step, y=DTU-SubmitAttempts]{plotdata/23.dat};
        \addlegendentry{GOAL-DTU}
        \addplot+[line width=2pt,mark=none,opacity=0.55] table [x=Step, y=VS-SubmitAttempts]{plotdata/23.dat};
        \addlegendentry{FIT BUT}
        \end{axis}
        \end{tikzpicture}
        }
        
        \caption{Submit: GOAL-DTU vs. FIT BUT (3)}
        \label{plot-fitbut-submit-3}
    \end{minipage}
\end{figure}

\begin{figure}[H]
    \centering
    \begin{minipage}{0.45\textwidth}
        \centering

        \resizebox{\textwidth}{!}{
        
        \begin{tikzpicture}
        \begin{axis}[
        xlabel=Steps,
        ylabel=Total tasks completed,
        xmin=0, xmax=500,
        ymin=0, ymax=20,
        xtick={0,100,...,500},
        ytick={0,5,...,20},
        ymajorgrids=true,
        xmajorgrids=true,
        grid style=dashed,
        ]
        \addplot+[line width=2pt,mark=none] table [x=Step, y=DTU-CompletedTasks]{plotdata/21.dat};
        \addlegendentry{GOAL-DTU}
        \addplot+[line width=2pt,mark=none,opacity=0.55] table [x=Step, y=VS-CompletedTasks]{plotdata/21.dat};
        \addlegendentry{FIT BUT}
        \end{axis}
        \end{tikzpicture}
        }

        \caption{Tasks: GOAL-DTU vs. FIT BUT (1)}
        \label{plot-fitbut-tasks-1}
    \end{minipage}\hfill
    \begin{minipage}{0.45\textwidth}
        \centering
        
        \resizebox{\textwidth}{!}{
        \begin{tikzpicture}
        \begin{axis}[
        xlabel=Steps,
        ylabel=Total tasks completed,
        xmin=0, xmax=500,
        ymin=0, ymax=20,
        xtick={0,100,...,500},
        ytick={0,5,...,20},
        ymajorgrids=true,
        xmajorgrids=true,
        grid style=dashed,
        ]
        \addplot+[line width=2pt,mark=none] table [x=Step, y=DTU-CompletedTasks]{plotdata/22.dat};
        \addlegendentry{GOAL-DTU}
        \addplot+[line width=2pt,mark=none,opacity=0.55] table [x=Step, y=VS-CompletedTasks]{plotdata/22.dat};
        \addlegendentry{FIT BUT}
        \end{axis}
        \end{tikzpicture}
        }
        
        \caption{Tasks: GOAL-DTU vs. FIT BUT (2)}
        \label{plot-fitbut-tasks-2}
    \end{minipage}
    
    \vspace{5.5em}
   
    \begin{minipage}{0.45\textwidth}
        \centering

        \resizebox{\textwidth}{!}{
        \begin{tikzpicture}
        \begin{axis}[
        xlabel=Steps,
        ylabel=Total tasks completed,
        xmin=0, xmax=500,
        ymin=0, ymax=20,
        xtick={0,100,...,500},
        ytick={0,5,...,20},
        ymajorgrids=true,
        xmajorgrids=true,
        grid style=dashed,
        ]
        \addplot+[line width=2pt,mark=none] table [x=Step, y=DTU-CompletedTasks]{plotdata/23.dat};
        \addlegendentry{GOAL-DTU}
        \addplot+[line width=2pt,mark=none,opacity=0.55] table [x=Step, y=VS-CompletedTasks]{plotdata/23.dat};
        \addlegendentry{FIT BUT}
        \end{axis}
        \end{tikzpicture}
        }

        \caption{Tasks: GOAL-DTU vs. FIT BUT (3)}
        \label{plot-fitbut-tasks-3}
    \end{minipage}\hfill
    \begin{minipage}{0.45\textwidth}
        \centering
        
        \resizebox{\textwidth}{!}{
        \begin{tikzpicture}
        \begin{axis}[
        xlabel=Steps,
        ylabel=Clear events,
        xmin=0, xmax=500,
        ymin=0, ymax=50,
        xtick={0,100,...,500},
        ytick={0,10,...,50},
        ymajorgrids=true,
        xmajorgrids=true,
        grid style=dashed,
        ]
        \addplot+[line width=2pt,mark=none] table [x=Step, y=ClearEvents]{plotdata/21.dat};
        \end{axis}
        \end{tikzpicture}
        }
        
        \caption{Clear: GOAL-DTU vs. FIT BUT (1)}
        \label{plot-fitbut-clear-1}
    \end{minipage}
    
    \vspace{5.5em}
   
    \begin{minipage}{0.45\textwidth}
        \centering

        \resizebox{\textwidth}{!}{
        
        \begin{tikzpicture}
        \begin{axis}[
        xlabel=Steps,
        ylabel=Clear events,
        xmin=0, xmax=500,
        ymin=0, ymax=50,
        xtick={0,100,...,500},
        ytick={0,10,...,50},
        ymajorgrids=true,
        xmajorgrids=true,
        grid style=dashed,
        ]
        \addplot+[line width=2pt,mark=none] table [x=Step, y=ClearEvents]{plotdata/22.dat};
        \end{axis}
        \end{tikzpicture}
        }

        \caption{Clear: GOAL-DTU vs. FIT BUT (2)}
        \label{plot-fitbut-clear-2}
    \end{minipage}\hfill
    \begin{minipage}{0.45\textwidth}
        \centering
        
        \resizebox{\textwidth}{!}{
        \begin{tikzpicture}
        \begin{axis}[
        xlabel=Steps,
        ylabel=Clear events,
        xmin=0, xmax=500,
        ymin=0, ymax=50,
        xtick={0,100,...,500},
        ytick={0,10,...,50},
        ymajorgrids=true,
        xmajorgrids=true,
        grid style=dashed,
        ]
        \addplot+[line width=2pt,mark=none] table [x=Step, y=ClearEvents]{plotdata/23.dat};
        \end{axis}
        \end{tikzpicture}
        }

        \caption{Clear: GOAL-DTU vs. FIT BUT (3)}
        \label{plot-fitbut-clear-3}
    \end{minipage}
\end{figure}

\begin{figure}[H]
    \centering
    \begin{minipage}{0.45\textwidth}
        \centering
        
        \resizebox{\textwidth}{!}{
        \begin{tikzpicture}
        \begin{axis}[
        xlabel=Steps,
        ylabel=Score,
        xmin=0, xmax=500,
        ymin=0, ymax=500,
        xtick={0,100,...,500},
        ytick={0,100,...,500},
        ymajorgrids=true,
        xmajorgrids=true,
        grid style=dashed,
        ]
        \addplot+[line width=2pt,mark=none] table [x=Step, y=DTU-Score]{plotdata/31.dat};
        \addlegendentry{GOAL-DTU}
        \addplot+[line width=2pt,mark=none,opacity=0.55] table [x=Step, y=VS-Score]{plotdata/31.dat};
        \addlegendentry{LFC}
        \end{axis}
        \end{tikzpicture}
        }
        
        \caption{Score: GOAL-DTU vs. LFC (1)}
        \label{plot-lfc-score-1}
    \end{minipage}\hfill
    \begin{minipage}{0.45\textwidth}
        \centering

        \resizebox{\textwidth}{!}{
        \begin{tikzpicture}
        \begin{axis}[
        xlabel=Steps,
        ylabel=Score,
        xmin=0, xmax=500,
        ymin=0, ymax=500,
        xtick={0,100,...,500},
        ytick={0,100,...,500},
        ymajorgrids=true,
        xmajorgrids=true,
        grid style=dashed,
        ]
        \addplot+[line width=2pt,mark=none] table [x=Step, y=DTU-Score]{plotdata/32.dat};
        \addlegendentry{GOAL-DTU}
        \addplot+[line width=2pt,mark=none,opacity=0.55] table [x=Step, y=VS-Score]{plotdata/32.dat};
        \addlegendentry{LFC}
        \end{axis}
        \end{tikzpicture}
        }

        \caption{Score: GOAL-DTU vs. LFC (2)}
        \label{plot-lfc-score-2}
    \end{minipage}
    
    \vspace{5.5em}
    
    \begin{minipage}{0.45\textwidth}
        \centering
        
        \resizebox{\textwidth}{!}{
        \begin{tikzpicture}
        \begin{axis}[
        xlabel=Steps,
        ylabel=Score,
        xmin=0, xmax=500,
        ymin=0, ymax=500,
        xtick={0,100,...,500},
        ytick={0,100,...,500},
        ymajorgrids=true,
        xmajorgrids=true,
        grid style=dashed,
        ]
        \addplot+[line width=2pt,mark=none] table [x=Step, y=DTU-Score]{plotdata/33.dat};
        \addlegendentry{GOAL-DTU}
        \addplot+[line width=2pt,mark=none,opacity=0.55] table [x=Step, y=VS-Score]{plotdata/33.dat};
        \addlegendentry{LFC}
        \end{axis}
        \end{tikzpicture}
        }
        
        \caption{Score: GOAL-DTU vs. LFC (3)}
        \label{plot-lfc-score-3}
    \end{minipage}\hfill
    \begin{minipage}{0.45\textwidth}
        \centering

        \resizebox{\textwidth}{!}{
        
        \begin{tikzpicture}
        \begin{axis}[
        xlabel=Steps,
        ylabel=Total blocks attached,
        xmin=0, xmax=500,
        ymin=0, ymax=150,
        xtick={0,100,...,500},
        ytick={0,25,...,150},
        ymajorgrids=true,
        xmajorgrids=true,
        grid style=dashed,
        ]
        \addplot+[line width=2pt,mark=none] table [x=Step, y=DTU-AttachedBlocks]{plotdata/31.dat};
        \addlegendentry{GOAL-DTU}
        \addplot+[line width=2pt,mark=none,opacity=0.55] table [x=Step, y=VS-AttachedBlocks]{plotdata/31.dat};
        \addlegendentry{LFC}
        \end{axis}
        \end{tikzpicture}
        }

        \caption{Blocks: GOAL-DTU vs. LFC (1)}
        \label{plot-lfc-blocks-1}
    \end{minipage}
    
    \vspace{5.5em}

    \begin{minipage}{0.45\textwidth}
        \centering
        
        \resizebox{\textwidth}{!}{
        \begin{tikzpicture}
        \begin{axis}[
        xlabel=Steps,
        ylabel=Total blocks attached,
        xmin=0, xmax=500,
        ymin=0, ymax=150,
        xtick={0,100,...,500},
        ytick={0,25,...,150},
        ymajorgrids=true,
        xmajorgrids=true,
        grid style=dashed,
        ]
        \addplot+[line width=2pt,mark=none] table [x=Step, y=DTU-AttachedBlocks]{plotdata/32.dat};
        \addlegendentry{GOAL-DTU}
        \addplot+[line width=2pt,mark=none,opacity=0.55] table [x=Step, y=VS-AttachedBlocks]{plotdata/32.dat};
        \addlegendentry{LFC}
        \end{axis}
        \end{tikzpicture}
        }
        
        \caption{Blocks: GOAL-DTU vs. LFC (2)}
        \label{plot-lfc-blocks-2}
    \end{minipage}\hfill
    \begin{minipage}{0.45\textwidth}
        \centering

        \resizebox{\textwidth}{!}{
        \begin{tikzpicture}
        \begin{axis}[
        xlabel=Steps,
        ylabel=Total blocks attached,
        xmin=0, xmax=500,
        ymin=0, ymax=150,
        xtick={0,100,...,500},
        ytick={0,25,...,150},
        ymajorgrids=true,
        xmajorgrids=true,
        grid style=dashed,
        ]
        \addplot+[line width=2pt,mark=none] table [x=Step, y=DTU-AttachedBlocks]{plotdata/33.dat};
        \addlegendentry{GOAL-DTU}
        \addplot+[line width=2pt,mark=none,opacity=0.55] table [x=Step, y=VS-AttachedBlocks]{plotdata/33.dat};
        \addlegendentry{LFC}
        \end{axis}
        \end{tikzpicture}
        }

        \caption{Blocks: GOAL-DTU vs. LFC (3)}
        \label{plot-lfc-blocks-3}
    \end{minipage}
\end{figure}

\begin{figure}[H]
    \centering
    \begin{minipage}{0.45\textwidth}
        \centering
        
        \resizebox{\textwidth}{!}{
        \begin{tikzpicture}
        \begin{axis}[
        xlabel=Steps,
        ylabel=Total submits attempted,
        xmin=0, xmax=500,
        ymin=0, ymax=10,
        xtick={0,100,...,500},
        ytick={0,1,...,10},
        ymajorgrids=true,
        xmajorgrids=true,
        grid style=dashed,
        ]
        \addplot+[line width=2pt,mark=none] table [x=Step, y=DTU-SubmitAttempts]{plotdata/31.dat};
        \addlegendentry{GOAL-DTU}
        \addplot+[line width=2pt,mark=none,opacity=0.55] table [x=Step, y=VS-SubmitAttempts]{plotdata/31.dat};
        \addlegendentry{LFC}
        \end{axis}
        \end{tikzpicture}
        }
        
        \caption{Submit: GOAL-DTU vs. LFC (1)}
        \label{plot-lfc-submit-1}
    \end{minipage}\hfill
    \begin{minipage}{0.45\textwidth}
        \centering

        \resizebox{\textwidth}{!}{
        \begin{tikzpicture}
        \begin{axis}[
        xlabel=Steps,
        ylabel=Total submits attempted,
        xmin=0, xmax=500,
        ymin=0, ymax=10,
        xtick={0,100,...,500},
        ytick={0,1,...,10},
        ymajorgrids=true,
        xmajorgrids=true,
        grid style=dashed,
        ]
        \addplot+[line width=2pt,mark=none] table [x=Step, y=DTU-SubmitAttempts]{plotdata/32.dat};
        \addlegendentry{GOAL-DTU}
        \addplot+[line width=2pt,mark=none,opacity=0.55] table [x=Step, y=VS-SubmitAttempts]{plotdata/32.dat};
        \addlegendentry{LFC}
        \end{axis}
        \end{tikzpicture}
        }

        \caption{Submit: GOAL-DTU vs. LFC (2)}
        \label{plot-lfc-submit-2}
    \end{minipage}
    
    \vspace{5.5em}
    
    \begin{minipage}{0.45\textwidth}
        \centering
        
        \resizebox{\textwidth}{!}{
        \begin{tikzpicture}
        \begin{axis}[
        xlabel=Steps,
        ylabel=Total submits attempted,
        xmin=0, xmax=500,
        ymin=0, ymax=10,
        xtick={0,100,...,500},
        ytick={0,1,...,10},
        ymajorgrids=true,
        xmajorgrids=true,
        grid style=dashed,
        ]
        \addplot+[line width=2pt,mark=none] table [x=Step, y=DTU-SubmitAttempts]{plotdata/33.dat};
        \addlegendentry{GOAL-DTU}
        \addplot+[line width=2pt,mark=none,opacity=0.55] table [x=Step, y=VS-SubmitAttempts]{plotdata/33.dat};
        \addlegendentry{LFC}
        \end{axis}
        \end{tikzpicture}
        }
        
        \caption{Submit: GOAL-DTU vs. LFC (3)}
        \label{plot-lfc-submit-3}
    \end{minipage}\hfill
    \begin{minipage}{0.45\textwidth}
        \centering

        \resizebox{\textwidth}{!}{
        
        \begin{tikzpicture}
        \begin{axis}[
        xlabel=Steps,
        ylabel=Total tasks completed,
        xmin=0, xmax=500,
        ymin=0, ymax=10,
        xtick={0,100,...,500},
        ytick={0,1,...,10},
        ymajorgrids=true,
        xmajorgrids=true,
        grid style=dashed,
        ]
        \addplot+[line width=2pt,mark=none] table [x=Step, y=DTU-CompletedTasks]{plotdata/31.dat};
        \addlegendentry{GOAL-DTU}
        \addplot+[line width=2pt,mark=none,opacity=0.55] table [x=Step, y=VS-CompletedTasks]{plotdata/31.dat};
        \addlegendentry{LFC}
        \end{axis}
        \end{tikzpicture}
        }

        \caption{Tasks: GOAL-DTU vs. LFC (1)}
        \label{plot-lfc-taks-1}
    \end{minipage}
    
    \vspace{5.5em}

    \begin{minipage}{0.45\textwidth}
        \centering
        
        \resizebox{\textwidth}{!}{
        \begin{tikzpicture}
        \begin{axis}[
        xlabel=Steps,
        ylabel=Total tasks completed,
        xmin=0, xmax=500,
        ymin=0, ymax=10,
        xtick={0,100,...,500},
        ytick={0,1,...,10},
        ymajorgrids=true,
        xmajorgrids=true,
        grid style=dashed,
        ]
        \addplot+[line width=2pt,mark=none] table [x=Step, y=DTU-CompletedTasks]{plotdata/32.dat};
        \addlegendentry{GOAL-DTU}
        \addplot+[line width=2pt,mark=none,opacity=0.55] table [x=Step, y=VS-CompletedTasks]{plotdata/32.dat};
        \addlegendentry{LFC}
        \end{axis}
        \end{tikzpicture}
        }
        
        \caption{Tasks: GOAL-DTU vs. LFC (2)}
        \label{plot-lfc-taks-2}
    \end{minipage}\hfill
    \begin{minipage}{0.45\textwidth}
        \centering

        \resizebox{\textwidth}{!}{
        \begin{tikzpicture}
        \begin{axis}[
        xlabel=Steps,
        ylabel=Total tasks completed,
        xmin=0, xmax=500,
        ymin=0, ymax=10,
        xtick={0,100,...,500},
        ytick={0,1,...,10},
        ymajorgrids=true,
        xmajorgrids=true,
        grid style=dashed,
        ]
        \addplot+[line width=2pt,mark=none] table [x=Step, y=DTU-CompletedTasks]{plotdata/33.dat};
        \addlegendentry{GOAL-DTU}
        \addplot+[line width=2pt,mark=none,opacity=0.55] table [x=Step, y=VS-CompletedTasks]{plotdata/33.dat};
        \addlegendentry{LFC}
        \end{axis}
        \end{tikzpicture}
        }

        \caption{Tasks: GOAL-DTU vs. LFC (3)}
        \label{plot-lfc-taks-3}
    \end{minipage}
\end{figure}

\begin{figure}[H]
    \centering
    \begin{minipage}{0.45\textwidth}
        \centering
        
        \resizebox{\textwidth}{!}{
        \begin{tikzpicture}
        \begin{axis}[
        xlabel=Steps,
        ylabel=Clear events,
        xmin=0, xmax=500,
        ymin=0, ymax=50,
        xtick={0,100,...,500},
        ytick={0,10,...,50},
        ymajorgrids=true,
        xmajorgrids=true,
        grid style=dashed,
        ]
        \addplot+[line width=2pt,mark=none] table [x=Step, y=ClearEvents]{plotdata/31.dat};
        \end{axis}
        \end{tikzpicture}
        }
        
        \caption{Clear: GOAL-DTU vs. LFC (1)}
        \label{plot-lfc-clear-1}
    \end{minipage}\hfill
    \begin{minipage}{0.45\textwidth}
        \centering

        \resizebox{\textwidth}{!}{
        
        \begin{tikzpicture}
        \begin{axis}[
        xlabel=Steps,
        ylabel=Clear events,
        xmin=0, xmax=500,
        ymin=0, ymax=50,
        xtick={0,100,...,500},
        ytick={0,10,...,50},
        ymajorgrids=true,
        xmajorgrids=true,
        grid style=dashed,
        ]
        \addplot+[line width=2pt,mark=none] table [x=Step, y=ClearEvents]{plotdata/32.dat};
        \end{axis}
        \end{tikzpicture}
        }

        \caption{Clear: GOAL-DTU vs. LFC (2)}
        \label{plot-lfc-clear-2}
    \end{minipage}
    
    \vspace{5.5em}

    \begin{minipage}{0.45\textwidth}
        \centering
        
        \resizebox{\textwidth}{!}{
        \begin{tikzpicture}
        \begin{axis}[
        xlabel=Steps,
        ylabel=Clear events,
        xmin=0, xmax=500,
        ymin=0, ymax=50,
        xtick={0,100,...,500},
        ytick={0,10,...,50},
        ymajorgrids=true,
        xmajorgrids=true,
        grid style=dashed,
        ]
        \addplot+[line width=2pt,mark=none] table [x=Step, y=ClearEvents]{plotdata/33.dat};
        \end{axis}
        \end{tikzpicture}
        }

        \caption{Clear: GOAL-DTU vs. LFC (3)}
        \label{plot-lfc-clear-3}
    \end{minipage}
\end{figure}

\newpage

\appendix

  \section{Team overview: short answers}

  \subsection{Participants and their background}
  \begin{description}
    \item \vskip0.5em\textbf{What was your motivation to participate in the contest?}
    To work on implementing a multi-agent system capable of competing in a realistic, albeit simulated, scenario.
    \item \vskip0.5em\textbf{What is the history of your group? (course project, thesis, $\ldots$)}

    The name of our team is GOAL-DTU. We participated in the contest in 2009 and 2010 as the Jason-DTU team \cite{Boss+2010,Vester+2011}, in 2011 and 2012 as the Python-DTU team \cite{Ettienne+2012,Villadsen+2013}, in 2013 and 2014 as the GOAL-DTU team \cite{Villadsen+2013-GOAL}, in 2015/2016 as the Python-DTU team \cite{Villadsen+2017} and in 2017 and 2018 as the Jason-DTU team \cite{Villadsen+2018}.

    The members of the team are as follows:
    \begin{itemize}
    \medskip
    \item J{\o}rgen Villadsen, PhD
    \medskip
    \item Alexander Birch Jensen, PhD student
    \medskip
    \end{itemize}

    Asta Halkj{\ae}r From, MSc student and now PhD student, was a consultant until the tournament started.

    We are affiliated with the Algorithms, Logic and Graphs section at DTU Compute, Department of Applied Mathematics and Computer Science, Technical University of Denmark (DTU).
    DTU Compute is located in the greater Copenhagen area.
    The main contact is associate professor J{\o}rgen Villadsen, email: \verb'jovi@dtu.dk'
    \item \vskip0.5em\textbf{What is your field of research? Which work therein is related?}
  We are responsible for the Artificial Intelligence and Algorithms study line of the MSc in Computer Science and Engineering programme.
  \end{description}

  \subsection{Statistics}
  \begin{description}
    \item \vskip0.5em\textbf{How much time did you invest in the contest (for programming, organizing your group, other)?}
    Approximately 200 man hours
    \item \vskip0.5em\textbf{How many lines of code did you produce for your final agent team?}
    Approximately 1000 lines
    \item \vskip0.5em\textbf{How many people were involved?}
    3 (1 programming)
    \item \vskip0.5em\textbf{When did you start working on your agents?}
    August 2019
  \end{description}

  \subsection{Agent system details}\label{sec:strategies}
  \begin{description}
    \item \vskip0.5em\textbf{How does the team work together? (i.e. coordination, information sharing, ...) How decentralised is your approach?}
    A task is delegated to a set of agents that are attached to the needed blocks. One agent is assigned as the so-called submit agent and the other agents follow/search for this submit agent before aligning the pattern in a goal area.
    Beyond this, each agent keeps track of the position of other agents. This information is exchanged when two agents are within vision range. The agents confirm their identify by agreeing on the part of the environment they both are able to perceive based on vision.
    \item \vskip0.5em\textbf{Do your agents make use of the following features: Planning, Learning, Organisations, Norms? If so, please elaborate briefly.}
    Planning is used when delegating tasks. Agents have set positions in the final pattern for submission.
    \item \vskip0.5em\textbf{Can your agents change their behavior during runtime? If so, what triggers the changes?}
    The agent's behavior changes if they are delegated a submission task. Furthermore, other agents will try to avoid blocking agents with a task.
    \item \vskip0.5em\textbf{Did you have to make changes to the team (e.g. fix critical bugs) during the contest?}
    We encountered timeout problems when the simulations ran too fast. We did not manage to resolve this beyond putting artificial limit. Furthermore, we did not manage to handle the automatic transition between simulations in each matchup. 
    \item \vskip0.5em\textbf{How did you go about debugging your system?}
    Partly using the debugger and partly using console output.
    \item \vskip0.5em\textbf{During the contest you were not allowed to watch the matches. How did you understand what your team of agents was doing? Did this understanding help you to improve your team's performance?}
    We tracked them using console output although this feature could be vastly improved. It did not help towards performance beyond discovering timeout problems in fast simulations.
    \item \vskip0.5em\textbf{Did you invest time in making your agents more robust? How?}
    Some robustness comes almost for free using GOAL as we never deeply commit to a plan. We also considered tracking if an agent ending being stuck, but ultimately the feature was not completed.
  \end{description}

  \subsection{Scenario and Strategy}
  \begin{description}
    \item \vskip0.5em\textbf{What is the main strategy of your agent team?}
    \begin{itemize}
      \item If the agent is selected to hand in blocks for a task (part of a task plan):
      \begin{itemize}
          \item Detach any attached blocks not needed for the task. The agent will only detach blocks if it considers it non-obstructive to future movement. If not, it will move until it reaches a position where it considers it safe to detach.
          \item Rotate the block into the position dictated by the task plan. If rotation is blocked, move until rotation is possible.
          \item If the agent observes part of the pattern to be handed in, or if the agent is the one to submit the task and is on a goal, wait for other agents (skip action).
          \item If the agent observes the entire pattern, connect with other agents as described by the task plan and then (the submit agent performs) submit.
          \item If the agent finds the submit agent (waiting in a goal area), move to place the attachment(s) as described by the task plan to form the final pattern.
          \item If the agent is the submit agent, move towards a goal area.
          \item If not the submit agent and believe that submit agent is in a goal area, move towards the position of the submit agent.
          \item If a goal area is known, move towards it (to see if we can find the submit agent there).
          \item Move into the most promising direction based on the exploration heuristics.
      \end{itemize}
      \item If the agent is not selected to hand in any task (not part of the current task plan)
      \begin{itemize}
          \item If a block or dispenser is in vision:
          \begin{itemize}
              \item Rotate such that a free attachment spot is facing the direction of the block/dispenser. If rotation is blocked, move.
              \item If it is a block, attach it to the agent.
              \item If it is a dispenser, request a block.
              \item If not next to the block, move towards it.
          \end{itemize}
          \item Move into the most promising direction based on the safe exploration heuristics.
      \end{itemize}
      \item Perform skip action.
  \end{itemize}

    \item \vskip0.5em\textbf{Your agents only got local perceptions of the whole scenario. Did your agents try to build a global view of the scenario for a specific purpose? If so, describe it briefly.}
    No global view is attempted beyond the position of other agents in the team.
    \item \vskip0.5em\textbf{How do your agents decide which tasks to complete?}
    Based on the currently collected blocks.
    \item \vskip0.5em\textbf{Do your agents form ad-hoc teams to complete a task?}
    Yes, see above.
    \item \vskip0.5em\textbf{Which aspect(s) of the scenario did you find particularly challenging?}
    The random map change events and deciding which blocks to clear (ultimately, we avoided trying to clear blocked paths).
    \item \vskip0.5em\textbf{If another developer needs to integrate your techniques into their code (i.e., same programming language tools), how easy is it to make that integration work?}
    That entirely depends on the programming language. Prolog is deeply integrated into much of the code.
  \end{description}

  \subsection{And the moral of it is \ldots}
  \begin{description}
    \item \vskip0.5em\textbf{What did you learn from participating in the contest?}
    We learned about using GOAL and general training in solving complex problems with no obvious solution.
    \item \vskip0.5em\textbf{What are the strong and weak points of your team?}
    Our agents are rather flexible and rarely idle. Weak points are that we are possibly too greedy collecting blocks which makes it harder to navigate the map as the simulation progresses.
    \item \vskip0.5em\textbf{Where did you benefit from your chosen programming language, methodology, tools, and algorithms?}
    GOAL helps our agents become flexible. We are forced to think in moment-to-moment reasoning and not just plans.
    \item \vskip0.5em\textbf{Which problems did you encounter because of your chosen technologies?}
    The freedom can make it harder to keep things simple as the complexity grows. Furthermore, GOAL had some integration issues with the provided EIS interface. We have to attempt changes to the source code to run.
    \item \vskip0.5em\textbf{Did you encounter new problems during the contest?}
    We were unaware of the feature that allows for multiple simulations without restarting. Furthermore, we had not tested GOAL with very fast simulations (the fact that we did not sent idle actions in our testing created an artificial slowdown).
    \item \vskip0.5em\textbf{Did playing against other agent teams bring about new insights on your own agents?}
    We learned that with another team playing the map became even harder to navigate based on our approach. However, we probably also won some matches by creating the same problem for the opponent.
    \item \vskip0.5em\textbf{What would you improve (wrt. your agents) if you wanted to participate in the same contest a week from now (or next year)?}
    Less rigid task submission plans and a less greedy approach to mindlessly collecting all blocks possible.
    \item \vskip0.5em\textbf{Which aspect of your team cost you the most time?}
    Navigating the map and trying to make the agents find each other for submission.
    \item \vskip0.5em\textbf{What can be improved regarding the contest/scenario for next year?}
    Set up test matches early using the contest setup to discover technical difficulties.
    \item \vskip0.5em\textbf{Why did your team perform as it did? Why did the other teams perform better/worse than you did?}
    We did not use roles for agents to help with different tasks. We saw other teams using interesting strategies to solve the tasks. Ultimately, we also had some false assumptions about the scenario which created artificial problems that could have been avoided. In the end, some parts of the design should be completely redone.
  \end{description}
 

\newpage

\begin{thebibliography}{1}

\bibitem{GOAL1}
Koen V. Hindriks, Frank S. de Boer, Wiebe van der Hoek, and John-Jules Ch. Meyer.
\newblock \emph{Agent Programming with Declarative Goals}.
\newblock Lecture Notes in Computer Science, 1986:228–243, Springer 2000.

\smallskip

\bibitem{GOAL2}
Koen V. Hindriks.
\newblock \emph{Programming Rational Agents in GOAL}.
\newblock Multi-Agent Programming, Languages, Tools and Applications, 119-157, Springer 2009.

\smallskip

\bibitem{GOAL3}
Koen V. Hindriks and J\"{u}rgen Dix:
\newblock \emph{GOAL: A Multi-Agent Programming Language Applied to an Exploration Game}. \newblock Agent-Oriented Software Engineering, 235-258, Springer 2014.

\smallskip

\bibitem{Boss+2010}
Niklas Skamriis Boss, Andreas Schmidt Jensen, and J{\o}rgen Villadsen.
\newblock \emph{Building Multi-Agent Systems Using Jason}.
\newblock Annals of Mathematics and Artificial Intelligence, 59:373-388, Springer 2010.

\smallskip

\bibitem{Vester+2011}
Steen Vester, Niklas Skamriis Boss, Andreas Schmidt Jensen, and J{\o}rgen Villadsen.
\newblock \emph{Improving Multi-Agent Systems Using Jason}.
\newblock Annals of Mathematics and Artificial Intelligence, 61:297-307, Springer 2011. 

\smallskip

\bibitem{Ettienne+2012}
Mikko Berggren Ettienne, Steen Vester, and J{\o}rgen Villadsen.
\newblock \emph{Implementing a Multi-Agent System in Python with an Auction-Based Agreement Approach}.
\newblock Lecture Notes in Computer Science, 7217:185-196, Springer 2012.

\smallskip

\bibitem{Villadsen+2013}
J{\o}rgen Villadsen, Andreas Schmidt Jensen, Mikko Berggren Ettienne, Steen Vester, Kenneth Balsiger Andersen, and Andreas Fr{\o}sig.
\newblock \emph{Reimplementing a Multi-Agent System in Python}.
\newblock Lecture Notes in Computer Science, 7837:205-216, Springer 2013.

\smallskip

\bibitem{Villadsen+2013-GOAL}
J{\o}rgen Villadsen, Andreas Schmidt Jensen, Nicolai Christian Christensen, Andreas Viktor Hess, Jannick Boese Johnsen, {\O}yvind Gr{\o}nland Woller, and Philip Bratt {\O}rum.
\newblock \emph{Engineering a Multi-Agent System in GOAL}.
\newblock Lecture Notes in Computer Science, 8245:329-338, Springer 2013.

\smallskip

\bibitem{Villadsen+2017}
J{\o}rgen Villadsen, Andreas~Halkj{\ae}r From, Salvador Jacobi and Nikolaj~N{\o}kkentved Larsen.
\newblock \emph{Multi-Agent Programming Contest 2016 --- The Python-DTU Team}.
\newblock International Journal of Agent-Oriented Software Engineering 6(1):86-100 2018.

\smallskip

\bibitem{Villadsen+2018}
J{\o}rgen Villadsen, Oliver Fleckenstein, Helge Hatteland and John Bruntse Larsen.
\newblock \emph{Engineering a Multi-Agent System in Jason and CArtAgO}.
\newblock Annals of Mathematics and Artificial Intelligence, 84:57-74, Springer 2018.

\end{thebibliography}
\end{document}